# Computer Aided Investigation: Visualization and Analysis of data from Mobile communication devices using Formal Concept Analysis.

Quist-Aphetsi Kester, MIEEE

Lecturer, Faculty of Informatics Ghana Technology University College, Accra, Ghana

Email: kquist-aphetsi@gtuc.edu.gh

*Abstract* - *There are challenges faced in today's world in terms of crime analysis when it comes to visualization and analysis of data in fighting crime. Activities of criminals in terms of who they communicate with, and how they are linked and related with other groups is very difficult to visualize and follow using traditional investigative procedures. In this world of terrorism, it is very important to know the network of individual suspects. It is also important to analyze the attributes of members of a network and the relationships that exist between them either directly or indirectly. This will make it easy for concepts to be built in aiding criminal investigations. However traditional approaches cannot be used to visualize and analyze data collected on individuals. With this current day where information systems play critical role in everyday life of every individual, it is easier to depend on digital information in fighting crime.*

*Effective computer tools and intelligent systems that are automated to analyze and interpret criminal data in real time effectively and efficiently are needed in fighting crime. These current computer systems should have the capability of providing intelligence from raw data and creating a visual graph which will make it easy for new concepts to be built and generated from crime data in order to solve understand and analyze crime patterns easily.*

*This paper proposes a new method of computer aided investigation by visualizing and analyzing data of mobile communication devices using Formal Concept Analysis, or Galois Lattices, a data analysis technique grounded on Lattice Theory and Propositional Calculus. This method considered the set of common and distinct attributes of data in such a way that categorizations are done based on related data with respect to time and events. This will help in building a more defined and conceptual systems for analysis of crime data that can easily be visualized and intelligently analyzed by computer systems.*

## I. Introduction

According to S Hinduja, many traditional crimes are now being aided or supported through the use of computers and networks, and wrongdoing previously never imagined has surfaced because of the incredible capabilities of information systems. Computer crimes are requiring law enforcement departments in general and criminal investigators in particular to tailor an increasing amount of their efforts toward successfully identifying, apprehending, and assisting in the successful prosecution of perpetrators. It is hoped that past knowledge can be assimilated with current observations of computer-related criminality to inform and guide the science of police investigations in the future. [1]

Virtually all societies in the modern world are troubled by criminal activities every day. Most of these activities are done by individuals or organized groups. While crime rates vary enormously from one country to another and from one region to another, criminal behavior remains a cause for concern amongst most members of the public. Solving crimes has been the prerogative of the criminal justice and law enforcement specialists. With the increasing use of the computerized systems to track crimes, computer data analysts have started helping the law enforcement officers and detectives to speed up the process of solving crimes. The most efficient and effective way of fighting crime today cannot be resourceful without geographical profiling. Criminal activities have become very complex in such a way that rapid monitory can only be achieved by using intelligent systems with geographical components. [2]

Criminal investigative approaches are needed to analyze connected series of crimes to determine the most probable area of offender residence. By incorporating both qualitative and quantitative methods, it assists in understanding spatial behavior of an offender and focusing the investigation to a smaller area of the community. Typically used in cases of serial murder or rape (but also arson, bombing, robbery, and other crimes), the technique helps police detectives prioritize information in large-scale major crime investigations that often involve hundreds or thousands of suspects and tips.[3]

A criminal pattern analysis is very crucial in combating crime. Computer systems have to be engaged in order to gather and interpret intelligence so as to control the criminal environment as well as influence effective decision making as in figure 1 below.

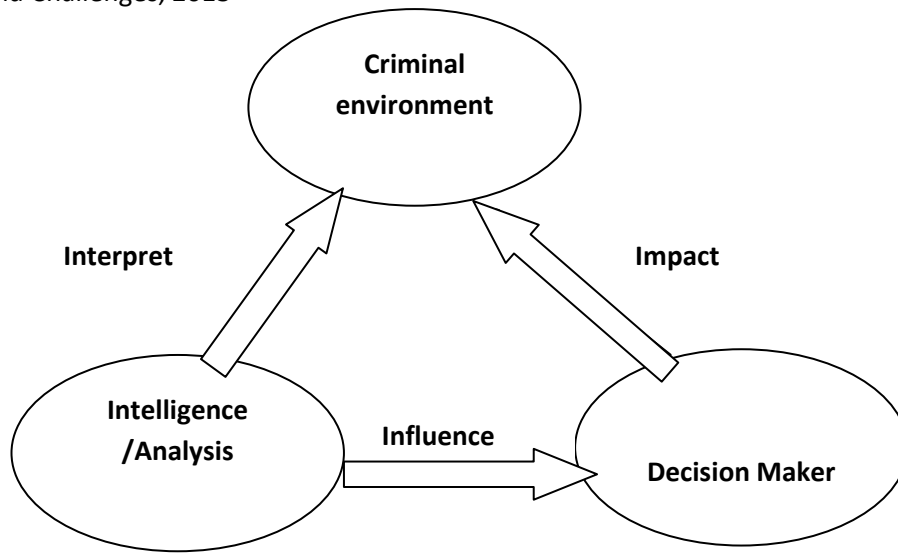

Figure 1: Pattern analysis theory

Formal Concept Analysis (FCA) is a mathematical theory of data analysis using formal contexts and concept lattices. It is a principled way of deriving a concept hierarchy or formal ontology from a collection of objects and their properties. Each concept in the hierarchy represents the set of objects sharing the same values for a certain set of properties; and each sub-concept in the hierarchy contains a subset of the objects in the concepts above it [11]. Its range of applications can be found in information and knowledge processing including visualization, data analysis and knowledge management.

This paper proposes a new method of computer aided investigation by visualizing and analyzing data of mobile communication devices using Formal Concept Analysis, or Galois Lattices, a data analysis technique grounded on Lattice Theory and Propositional Calculus. This method considered the set of common and distinct attributes of data in such a way that categorizations are done based on related data with respect to time and events. This will help in building a more defined and conceptual systems for analysis of crime data that can easily be visualized and intelligently analyzed by computer systems.

The organization of this paper is as follows, section II of this paper provides details of the related work in the domain of crime pattern analysis and application of FCA in the field of computer science. Section III proposes how the FCA will be used to classify and analyze data with respect to events that has to do with crime. Section IV provides application and results: provides the use of FCA in crime analysis. The last section of this paper, section V, Concludes the paper.

## II. Literature Review

Crime activities are geospatial phenomena and as such are geospatially, thematically and temporally correlated. Thus, crime datasets must be interpreted and analyzed in conjunction with various factors that can contribute to the formulation of crime. Discovering these correlations allows a deeper insight into the complex nature of criminal behavior.

There are challenges faced in today's world in terms of crime analysis when it comes to graphical visualization of crime patterns. Geographical representation of crime scenes and crime types become very important in gathering intelligence about crimes. This provides a very dynamic and easy way of monitoring criminal activities and analyzing them as well as producing effective countermeasures and preventive measures in solving them. (QA Kester 2013) proposed a new method of visualizing and analyzing crime patterns based on geographical crime data.[2]

In the United States , Federals collaborate with local law enforcement and prosecutors to share intelligence and efforts through teamwork has demonstrated effectiveness in addressing traditional crimes involving drugs, weapons, gangs, and violence (McGarrell & Schlegel, 1993; Russell-Einhorn, 2004)[4][5]. By extension, many scholars and practitioners have asserted the importance of forming comparable teams to combat computer crime with the hope of similar positive outcomes (see e.g., Conly & McEwen, 1990)[6].[1]

P. Rogerson and Y. Sun described a new procedure for detecting changes over time in the spatial pattern of point events, combining the nearest neighbor statistic and cumulative sum methods. The method results in the rapid detection of deviations from expected geographic patterns. The method was illustrated using 1996 arson data from the Buffalo, NY, Police Department. [7]

The appearance of patterns could be found in different modalities of a domain, where the different modalities refer to the data sources that constitute different aspects of a domain. Particularly, the domain that refers to crime and the different modalities refer to the different data sources within the crime domain such as offender data, weapon data, etc. In addition, patterns also exist in different levels of granularity for each modality. In order to have a thorough understanding a domain, it is important to reveal hidden patterns through the data explorations at different levels of granularity and for each modality. Therefore, Yee Ling Boo and Alahakoon, D presented a new model for identifying patterns that exist in different levels of granularity for different modes of crime data. A hierarchical clustering approach - growing self organizing maps (GSOM) has been deployed. The model was further enhanced with experiments that exhibit the significance of exploring data at different granularities. [8]

Formal concept analysis (FCA) is a method of data analysis with growing popularity across various domains .FCA analyzes data which describe relationship between a particular set of objects and a particular set of attributes. Such data commonly appear in many areas of human activities. FCA produces two kinds of output from the input data. The first is a concept lattice. A concept lattice is a collection of formal concepts in the data which are hierarchically ordered by a sub-concept super-concept relation. Formal concepts are particular clusters which represent natural human-like concepts such as "organism living in water", "car with all wheel drive system", "number divisible by 3 and 4",etc. The second output of FCA is a collection of so-called attribute implications. An attribute implication describes a particular dependency which is valid in the data such as "every number divisible by 3 and 4 is divisible by 6","every respondent with age over 60 is retired", etc.[9]

Modern police organizations and intelligence services are adopting the use of FCA in crime pattern analysis for tracking down criminal suspects through the integration of heterogeneous data sources and visualizing this information so that a human expert can gain insight in the data [19].

### III. Methodology

Formal Concept Analysis, or Galois Lattices, is a data analysis technique grounded on Lattice Theory and Propositional Calculus. This paper proposes a new method based on FCA method and considered the set of common and distinct attributes of data collected from mobile communication devices of suspects in such a way that categorizations are done based on related data with respect to time and events.

In FCA a formal context consists of a set of objects, *G*, a set of attributes, *M*, and a relation between *G* and *M*, $I \subseteq G \times M$. A formal concept is a pair *(A, B)* where $A \subseteq G$ and $B \subseteq M$. Every object in *A* has every attribute in *B*. For every object in *G* that is not in *A*, there is an attribute in *B* that that object does not have. For every attribute in *M* that is not in *B* there is an object in *A* that does not have that attribute. *A* is called the extent of the concept and *B* is called the intent of the concept.

If $g \in A$ and $m \in B$ then $(g,m) \in I$, or *gIm*.

A formal context is a triple *(G, M, I)*, where

- *G* is a set of objects,

- *M* is a set of attributes

- and *I* is a relation between *G* and *M*.

- $(g,m) \in I$ is read as "object g has attribute m".

For $A \subseteq G$, we define

$A' := \{m \in M \mid \forall g \in A : (g,m) \in I \}$.

For $B \subseteq M$, we define dually

$B' := \{g \in G \mid \forall m \in B : (g,m) \in I \}$.

For *A, A1, A2* $\subseteq G$ holds:

- $A1 \subseteq A2 \Rightarrow A'2 \subseteq A'1$

- $A1 \subseteq A''$

- $A' = A'''$

For *B, B1, B2* $\subseteq M$ holds:

- $B1 \subseteq B2 \Rightarrow B'2 \subseteq B'1$

- $B \subseteq B''$

- $B` = B```$

A formal concept is a pair *(A, B)* where

- *A* is a set of objects (the extent of the concept),

- *B* is a set of attributes (the intent of the concept),

- $A` = B$ and $B` = A$.

The concept lattice of a formal context *(G, M, I)* is the set of all formal concepts of *(G, M, I)*, together with the partial order

$(A1, B1) \leq (A2, B2): \Leftrightarrow A1 \subseteq A2 \ (\Leftrightarrow B1 \supseteq B2)$.

The concept lattice is denoted by *(G,M,I)* .

- Theorem: The concept lattice is a lattice, i.e. for two concepts *(A1, B1)* and *(A2, B2)*, there is always

- a greatest common subconcept: $(A1 \cap A2, (B1 \cup B2)``)$

- and a least common superconcept: $((A1 \cup A2)``, B1 \cap B2)$

More general, it is even a complete lattice, i.e. the greatest common subconcept and the least common superconcept exist for all (finite and infinite) sets of concepts.

Corollary: The set of all concept intents of a formal context is a closure system. The corresponding closure operator is $h(X) := X``$.

An implication *X→Y* holds in a context, if every object having all attributes in *X* also has all attributes in *Y*.

Def.: Let $X \subseteq M$. The attributes in *X* are independent, if there are no trivial dependencies between them.

## IV. The Applications and Results

We consider crime events such that $E$ is the set of crime events and $E_1, E_2...E_n$ are subsets of $E$. Let E= {$E_1, E_2...E_n$} = {murder, robbery, drugs, assault, burglary, Money Laundering, Bombing, Theft, Rape, Kidnapping …..}. Let the objects be the events. Let the geographical locations in which the events occurred at be *(A, B, C, D, E, F, G, H, I, J and K)* as indicated on the map below in figure 2. Let P be the set of suspects intended to be investigated and $P_1, P_2, P_3$ ….. $P_n$ are the elements that belong to *P*.

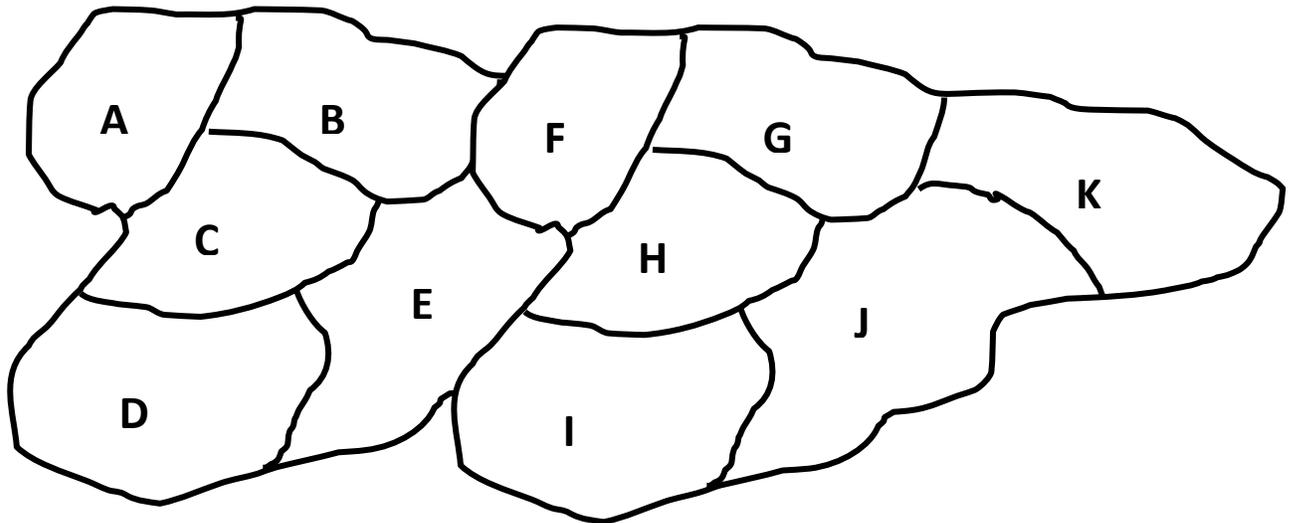

Figure 2: Geographical locations of a map

Table 1: Events x (Geographical locations and Persons)

| | Geographical Locations | | | | | | | | | | Persons | | | | | | | | | |
|---|---|---|---|---|---|---|---|---|---|---|---|---|---|---|---|---|---|---|---|---|
| Events | a | b | c | d | e | f | g | h | i | j | k | P1 | P2 | P4 | P5 | P6 | P7 | P8 | P3 | P9 | P10 |
| 1 | X | | X | X | | X | | X | | X | X | X | | | X | X | | | | | |
| 2 | | X | | | | | | | | | | | X | | X | | | | X | | X | X |
| 3 | X | | X | X | | X | X | | X | X | | | | | | X | X | | X | | |
| 4 | | X | | | | | X | | | | | | | X | | | X | | | X | |
| 5 | | X | | | X | X | | X | | X | | X | | | X | X | | | X | | X |
| 6 | X | | | | | | | X | | X | | | X | | | | | | | X | |
| 7 | | | X | X | | | X | | | | | | | | | X | | X | X | | |
| 8 | X | | | | X | | | X | X | | | X | | | | | | X | X | X | X |
| 9 | | X | | | X | | X | | | X | | | | | | X | X | X | | | X |
| 10 | | | | | | | | | | | X | | | X | X | | | | | | |

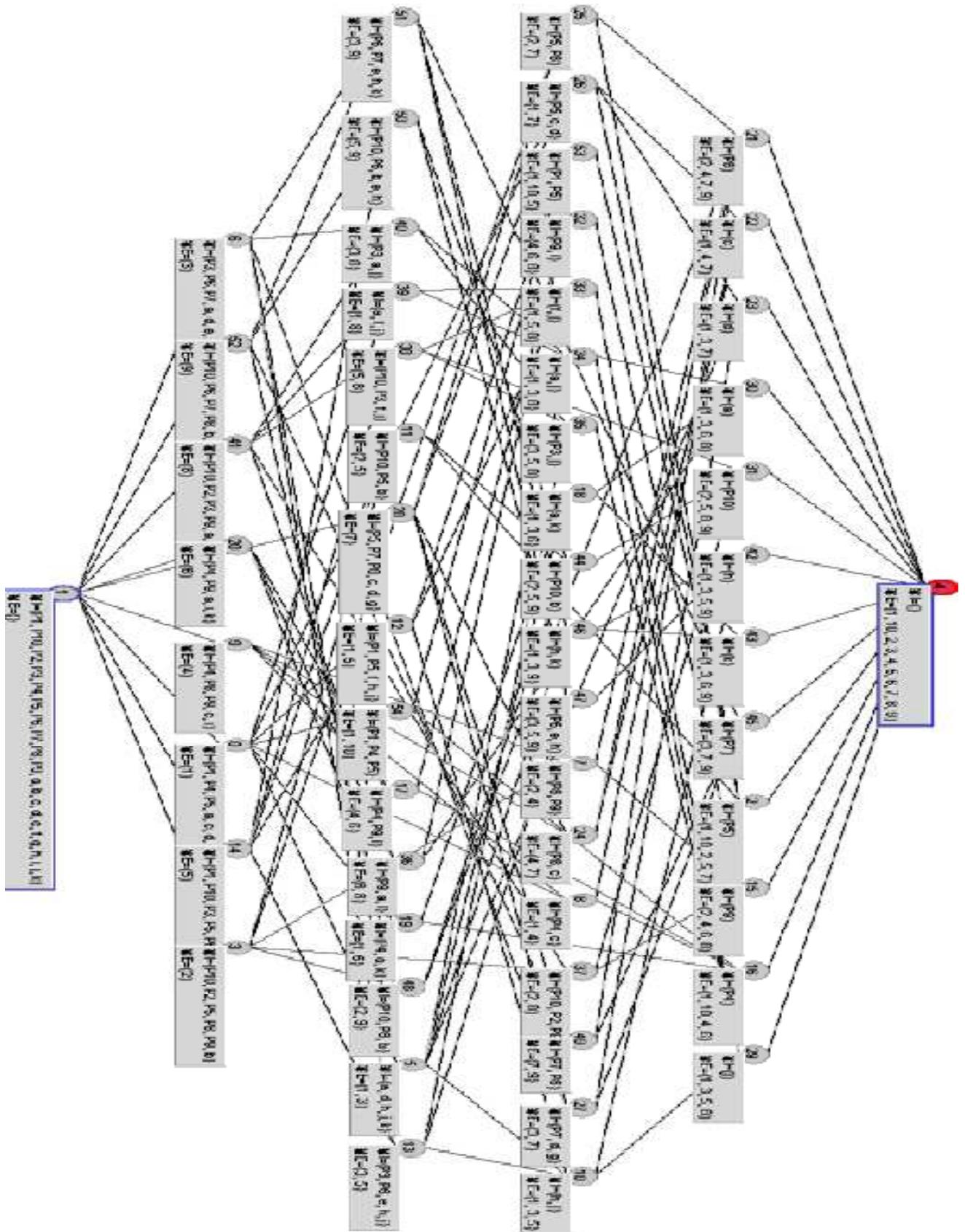

Figure 3: Galois lattices of intents and extents

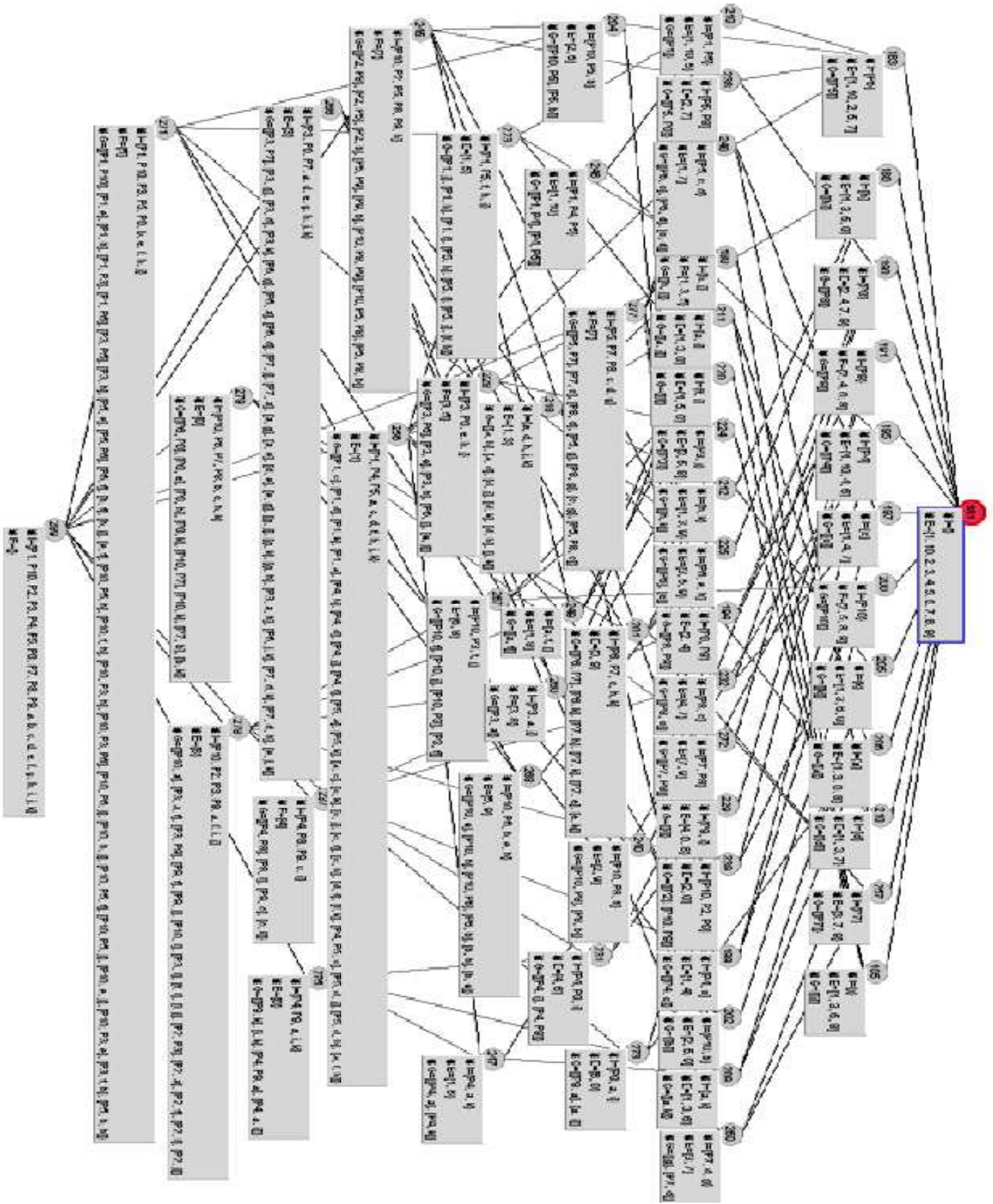

Figure 4: Galois lattice of *(G, M, I)*

The concept lattice of a formal context *(G, M, I)* as shown in figure 4, is the set of all formal concepts of *(G, M, I)*, together with the partial order *(A1, B1) ≤ (A2, B2):* ⇔ *A1 ⊆ A2 (*⇔ *B1 ⊇ B2)*.

It is clearly seen visually from figure 3 that the intents of concepts *24, 25, 26, 27* and *49* are subsumed by the intents of concept *28*. This means that all suspects involved in events *24, 25, 26, 27* and *49*, have a higher probability of committing crime at geographical location *c, d,* and *g* of concept *28*. Relationally and based on figure 4, the deduction is accurate.

From figure 4, concept *277* we have intent *I= {P5, P7, P8, c, d, g}*, extent *E= {7}* and set of attributes *G= {[P5, P7], [P7, c], [P8, g], [c, g], [P5, P8, c]}*. We can draw several conclusions from the analysis such as the persons involved in event of *E= {7}* are likely to have been working in a network group and events at places c, d and g are likely to have a link also with event of *E= {7}*.

### V. Conclusion

A formal concept analysis was used to analyze crime data based on information gathered from suspects' mobile communication devices such as mobile phone, tablets etc. Visualization of relationships between the occurrences of various crime events within different geographical areas was achieved successfully. This method considered the set of common and distinct attributes of data in such a way that categorization was done based on relationships between concepts. The result from the approach will help in building a more defined and conceptual systems that will make data relationships to be easily visualized and intelligently analyzed by computer aided investigation systems.